\documentclass[aps,pre,endfloats, preprint, amsmath,amssymb,showpacs,showkeywords,groupedaddress]{revtex4}
\usepackage{dcolumn}
\usepackage{bm}
\usepackage[dvips]{graphicx}

\def \floatp@sw{\true@sw}
\makeatletter
\def\@dotsep{4.5}
\makeatother

\begin{document}

\title{A maximum entropy model for opinions in social groups}

\author{Sergio Davis}
\email{sdavis@gnm.cl}

\author{Yasm\'{\i}n Navarrete}
\email{yasmin@gnm.cl}

\author{Gonzalo Guti\'errez}
\email{gonzalo@fisica.ciencias.uchile.cl}

\affiliation{Grupo de Nanomateriales, Departamento de F\'{i}sica, Facultad de
Ciencias, Universidad de Chile, Casilla 653, Santiago, Chile}

\date{\today}

\begin{abstract}
We study how the opinions of a group of individuals determine their spatial
distribution and connectivity, through an agent-based model. The interaction between 
agents is described by a Hamiltonian in which agents are allowed to move freely 
without an underlying lattice (the average network topology connecting them is determined from the
parameters). This kind of model was derived using maximum entropy statistical inference 
under fixed expectation values of certain probabilities that (we propose) are relevant to social
organization. Control parameters emerge as Lagrange multipliers of the maximum
entropy problem, and they can be associated with the level of consequence between 
the personal beliefs and external opinions, and the tendency to socialize with
peers of similar or opposing views. These parameters define a phase diagram for the social system,
which we studied using Monte Carlo Metropolis simulations. Our model presents both first and 
second-order phase transitions, depending on the ratio between the internal consequence and 
the interaction with others. We have found a critical value for the level of internal consequence, 
below which the personal beliefs of the agents seem to be irrelevant.
\end{abstract}



\maketitle

\section{Introduction}

Agent-based models have been extensively used to comprehend some aspects of social
behavior. In these models, an individual is typically represented either as a (free) particle 
or as a node in a network with a given topology, to which there is some additional
information attached (its inner degrees of freedom). It is interesting to note
that there is some overlap with physical models, and this promises to bring new insight to 
questions such as the possible processes of formation of social groups, what
determines their stability and their evolution, in terms of ``microscopical
details'' such as the kinds of interaction among individuals\cite{Ball2002,Castellano2009}. 
Two models in particular have gathered considerable interest and have stood the test of time, these are the 
Sznajd model for the dynamics of opinions and consensus in a population~\cite{Sznajd2000,Sabatelli2004,Wang2008}, 
and Axelrod's model for the dissemination of
culture~\cite{Axelrod1997,Castellano2000}. Unfortunately, obtaining analytical
results from these (and other, similar) models is sometimes a daunting task, so
one has to resort to numerical simulations. 

Several approaches have been devised for the numerical treatment of agent-based models. Among them we 
find non-deterministic cellular automata~\cite{Kacperski1999}, variants of the Potts model~\cite{Liu2001,Schulze2005}, 
use of the Langevin equation for the description of stochastic dynamics~\cite{Hu2009}, as well as Hamiltonian 
models~\cite{Fronczak2006} together with other tools coming from Statistical Mechanics. Another, 
completely different line is the use of empirical models in analogy with physical phenomena~\cite{Ausloos2007,Ausloos2009}.

Agent-based models present several interesting properties, such as the emergence of different kinds 
of phase transitions~\cite{Castellano2000,Klemm2003,Klemm2003b,Fronczak2007}. In these phases we can recognize 
different regimes (ordered and disordered equilibrium states, metastable states) which
appeal to our intuitions about social phenomena. Interestingly, recent
studies~\cite{Biely2008,Kozma2008,Benczik2009} have drifted away from fixed-lattice models towards adaptive networks, 
taking into account the social mobility of individuals.

The central motivation for the present paper has been the development of an agent-based social model of 
opinion that explicitly incorporates the mobility of individuals in a coherent and least-biased manner.
In order to achieve this goal, we have recognized the existence of two important factors which are relevant 
to the social dynamics: the level of consequence of individuals with respect to their private, internal opinions 
on a given issue, and the strength and type of coupling between individuals. This coupling can be assortative (meaning 
the individual tends to be close only to those with the same opinion) or disassortative (where the tendency is the opposite, 
i.e., to be surrounded only by those with different opinions to their own).

Focusing on just these two aspects of the social phenomena (internal consequence and interaction 
with others) we have followed the formalism of maximum entropy statistical inference~\cite{Jaynes1957} 
to construct an unbiased probability distribution for the state of our system under known degrees of 
consequence and coupling. From this probability distribution we deduced the corresponding objective function or 
Hamiltonian, which the system minimizes in the absence of noise. In the definition of our model, we aim to assume 
nothing besides the observed qualities of interaction and consequence in agents.
The Hamiltonian function by itself provides a complete description of the equilibrium states, including the spatial 
distribution of agents, the level of homogeneity or diversity of opinions within a society, among other properties.
The resulting Hamiltonian resembles the well-known Potts model~\cite{Wu1982}, and is actually a generalization of it, 
in which the topology of the lattice is no longer fixed, and where an additional term in the form of a local field is 
considered. This term introduces a tendency of an individual to resist the influence of its neighbors. The fact that our 
model lifts the restriction of a fixed lattice (on which agents traditionally ``live'') grants the agents a complete 
freedom of movement and minimizes the number of assumptions.

This paper is organized as follows. In section II we motivate our constraints
and derive our model using the maximum entropy formalism. Section III comments
on the Hamiltonian thus derived and its resemblance to the Potts Hamiltonian. In
section IV we present analytical results as well as Monte Carlo simulations of
the phase diagram for our model. Finally we provide some concluding remarks. 

\section{Derivation of the model using statistical inference}

Suppose the following. For a given issue there are $Q$ alternative positions or possible 
opinions an individual can adopt as his/her own. For instance, these could be political
views to adhere to, sports teams to support, musical styles people enjoy, and so
on. A particular individual could have a preferred choice, which is private to
him/her, and could express the same choice, or a different one, to others,
for instance depending on social pressure (to ``fit in''). For example,
consider a political election with just three parties: left-wing, right-wing and
independents. An individual with, say, independent views could instead
express right-wing views in order to be socially accepted in an
exclusive club reunion among mostly right-wing peers. It is natural to think also that some 
individuals will prefer the company of like-minded peers, sharing the same preferred opinions, 
while others will seek the company of people with postures different from their own.

In order to construct our mathematical model we will consider $N$ agents
(labelled by an index $i=1,\ldots,N$), each one having the following attributes:
\begin{enumerate}
\item [(a)] a continuous position vector $\mathbf{r}_i$ in a two-dimensional space,
\item [(b)] a personal or internal opinion (or belief) $B_i$, which represents
what the agent ``really believes'', and is encoded as an integer between 1 and $Q$, and
\item [(c)] an external opinion $S_i$, which represents what the agent actually
expresses in front of others (and is also encoded as an integer between 1 and $Q$). 
\end{enumerate}

In the example above, the individual with independent views has $B=$3 and $S=2$. In a social context, 
definitions (a), (b) and (c) encode the information we have about individuals being free to move around, 
holding a (sometimes strong) belief about a particular issue and expressing an
opinion about said issue. The opinions represented may not represent faithfully
what the individuals really believe, and in such cases $S_i$ $\neq$ $B_i$. 

We will assume that the set of internal beliefs $B_i$ of the agents are fixed
(``quenched''), so that the coordinates of a given agent are only its position and
external opinion $S_i$. In other words, we are studying the behavior of a
particular set of individuals, within a time interval short enough so that none
of them has the opportunity to change its belief. Of course, to infer the
behavior of a generic population, we would need to marginalize (average over)
the belief variables $B_i$ under a given statistical model $P(B_1, \ldots,
B_N|H)$, for instance, a multinomial distribution.

On the other hand, we will not assume any particular topology or interaction between 
agents at this point, those will be provided by the maximum entropy inference procedure. 
However, we will assume that there is some ``locality'' to the influence in terms of opinions, 
that is, there is some measure of distance among agents such that 
a particular agent is only concerned with the opinions of peers closer than a given radius $R_c$.

Now, in order to arrive at a well defined Hamiltonian, we need to introduce the main
assumptions of the model. The first assumption we will make, besides the existence of all the attributes
described above (definitions (a), (b) and (c)), is that the phenomenon of consequence is completely described by the 
probability of an agent being consequent, i.e., the probability that its internal and external opinion actually agree. 
This probability can be written in Bayesian notation as $P(S_i = B_i|H)$ (the $H$ in the conditional is the generic Bayesian ``placeholder'' for the current
context or state of information) and we will denote it shortly by $P_C$.

\begin{figure}
\includegraphics[scale=0.25]{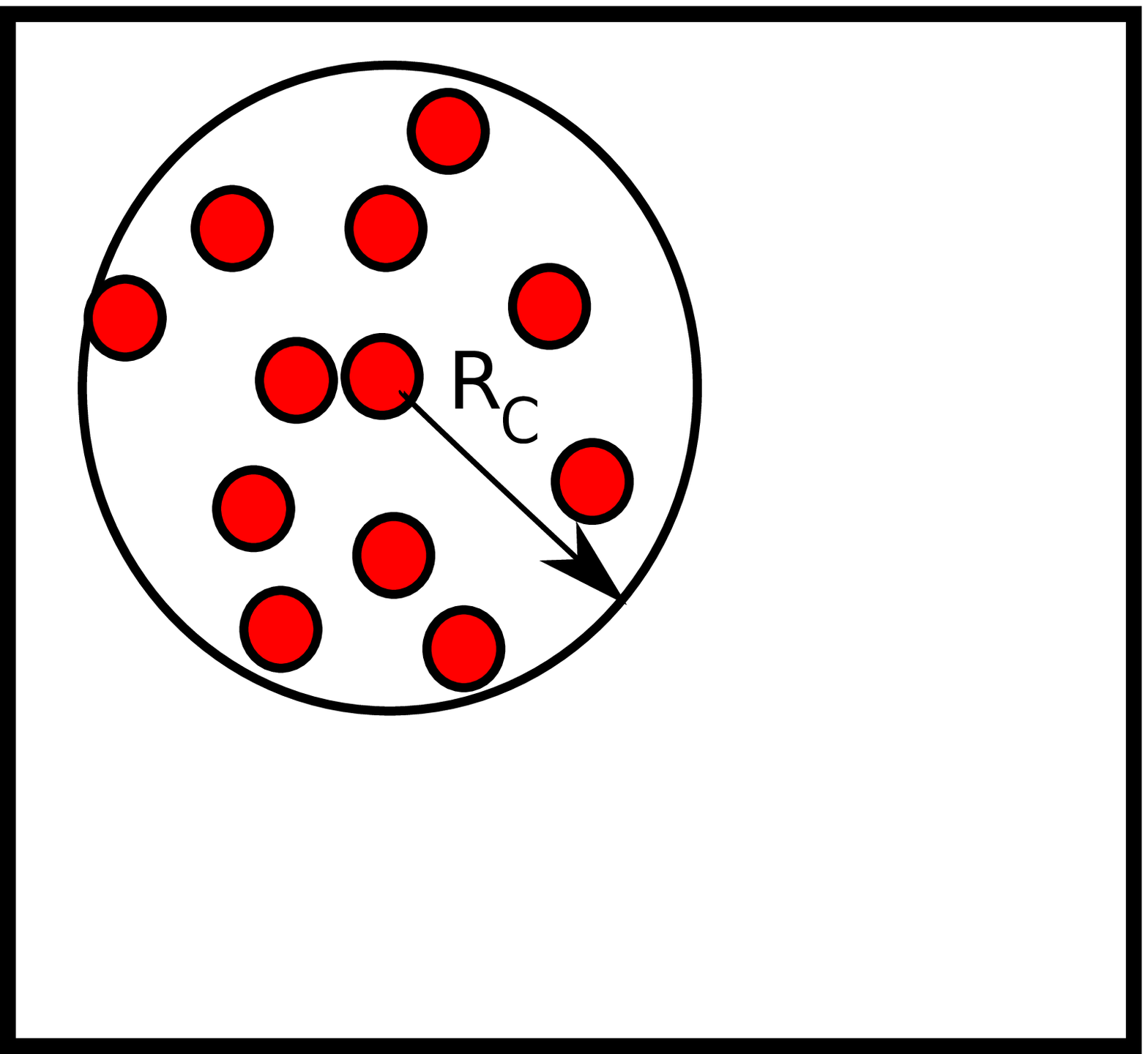}
\includegraphics[scale=0.25]{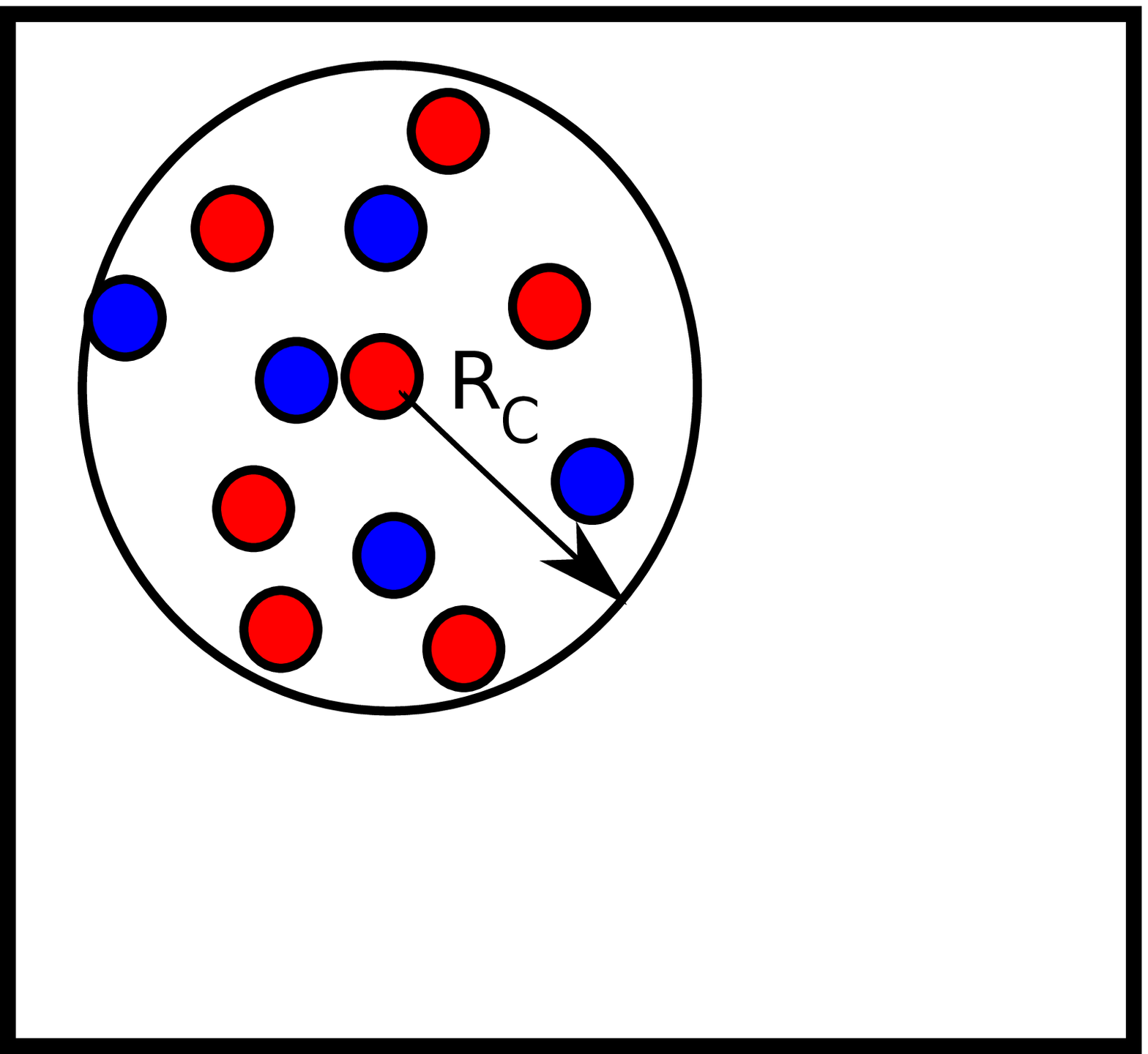}
\caption{A schematic diagram depicting two states of a system for which we could
attempt to infer an assortative association. The left panel is consistent with
assortativity, but only until we see the right panel we realize it is only an
effect of a tendency for closeness between agents (regardless of their color).}
\label{fig_closeness}
\end{figure}

Our second assumption is that the joint probability of agreement and closeness, i.e., that two agents simultaneously 
agree in opinion ($S_i=S_j$) and are spatially close ($r_{ij} < R_c$), written by $P(S_i = S_j \wedge r_{ij} < R_c|H)$ and
denoted by $P_J$, is relevant to the phenomenon of assortativity. However, this
by itself is not sufficient: for instance, observing a situation where both
agreement and closeness is common (Fig. \ref{fig_closeness}, left panel) does
not imply the existence of correlation between closeness and agreement. We could
also observe situations like the right panel of Fig. \ref{fig_closeness}, and
then we infer that in this particular case a tendency to closeness explains everything. 

Therefore we will include as a third assumption, the relevance of the
probability of closeness, namely $P(r_{ij} < R_c|H)=P_R$, to a description of
the phenomenon of assortative/disassortative association.

We can write all three probabilities as constraints in the form of known expectation values, 

\begin{eqnarray}
P(S_i = B_i|H) = P_C = \frac{1}{N}\Big<\sum_{i=1}^N \delta(S_i,B_i)\Big> \\
P(S_i = S_j \wedge r_{ij} < R_c|H) = \frac{1}{N(N-1)}\Big<\sum_{i=1}\sum_{j\neq
i}\Theta(R_c-r_{ij})\delta(S_i, S_j)\Big> \\
P(r_{ij} < R_c|H) = \frac{1}{N(N-1)}\Big<\sum_{i=1}\sum_{j\neq
i}\Theta(R_c-r_{ij})\Big> 
\label{eq_P}
\end{eqnarray}
where $\delta(a, b)$ is Kronecker's delta function and $\Theta(x)$ is Heaviside's step function. 
Here we can outline the problem as a case of maximum entropy
(MaxEnt) inference~\cite{Jaynes1957}. The most unbiased estimation of the full probability of being
in state $\Gamma$=\{$S_1$,$S_2$,\ldots,$S_N$,$\mathbf{r}_1$,$\mathbf{r}_2$,\ldots,$\mathbf{r}_N$\} is the one that 
maximizes Shannon's information entropy

\begin{equation}
\mathcal{S} = -\sum_{S_1,\ldots,S_N} \int d\mathbf{r}_1\mathbf{r}_2\ldots\mathbf{r}_N
P(\Gamma|H) \log_2 \frac{P(\Gamma|H)}{P(\Gamma|H_0)}.
\end{equation}
subjected to the known expectation values at the right-hand side of Eq.
\ref{eq_P}. Note that, as the $\mathbf{r}_i$ are continuous we have to include
$P(\Gamma|H_0)$ as an invariant measure~\cite{Jaynes2003} for the configuration
space $\Gamma$. We will consider it a constant which does not influence the maximization procedure. 
The maximum entropy solution for $P(\Gamma|H)$ has the following form,

\begin{equation}
\ln P(\Gamma|H) = -\ln Z + \frac{\lambda_C}{N}\sum_{i=1}^N\delta(S_i, B_i) +
\frac{1}{N(N-1)}\sum_{i=1}^N\sum_{j\neq i}\Theta(R_c-r_{ij})(\lambda_J\delta(S_i, S_j)+\lambda_R).
\end{equation}

which, by defining a Hamiltonian function $\mathcal{H}$ (in analogy with statistical
mechanics), as 

\begin{equation}
\beta\mathcal{H} = -\frac{\lambda_C}{N}\sum_{i=1}^N\delta(S_i, B_i) -
\frac{1}{N(N-1)}\sum_{i=1}^N\sum_{j\neq i}\Theta(R_c-r_{ij})(\lambda_J\delta(S_i, S_j)+\lambda_R).
\label{eq_hamiltonian}
\end{equation}
can be put in the form

\begin{equation}
P(\Gamma|H) = \frac{1}{Z}\exp(-\beta\mathcal{H}(\Gamma))
\label{eq_probability}
\end{equation}
where $Z=\sum_{S_1,\ldots,S_N}\int d\mathbf{r}_1\mathbf{r}_2\ldots\mathbf{r}_N \exp(-\beta\mathcal{H}(\Gamma))$ 
is the so-called partition function. 

Here $\lambda_C$, $\lambda_J$ and $\lambda_R$ are the Lagrange multipliers included to solve the 
variational problem, which are obtained from the constraints themselves,

\begin{equation}
P((S_i = S_j) \wedge (r_{ij} < R_c)|H) = -\frac{\partial}{\partial \lambda_J}\ln
Z(\lambda_J, \lambda_C, \lambda_R) \\
\label{eq_multi_PJ}
\end{equation}
\begin{equation}
P(S_i = B_i|H) = -\frac{\partial}{\partial \lambda_C}\ln Z(\lambda_J, \lambda_C,
\lambda_R) \\
\label{eq_multi_PC}
\end{equation}
\begin{equation}
P(r_{ij} < R_c|H) = -\frac{\partial}{\partial \lambda_R}\ln Z(\lambda_J,
\lambda_C, \lambda_R).
\label{eq_multi_PR}
\end{equation}

Eq. \ref{eq_probability} is completely equivalent to a canonical ensemble, and $\mathcal{H}$ is akin to a
Hamiltonian or energy function for our system, which emerges from the combination of constraints imposed in the MaxEnt 
inference. Note that $\beta$ is not an independent Lagrange multiplier, it is just an arbitrary factor fixing the scale
of $\mathcal{H}$, and can be interpreted in just the same way as the inverse temperature in Statistical Mechanics.
In fact, if we regard the Hamiltonian as an energy function, we have the usual relationship,

\begin{equation}
-\frac{\partial}{\partial \beta}\ln Z = \Big<\mathcal{H}\Big> = E(\beta).
\label{eq_beta}
\end{equation}

In the following, we will call $T=1/\beta$ the \emph{social temperature} to
preserve this useful analogy. It is important to note that, in our case, as the
Hamiltonian in Eq. \ref{eq_hamiltonian} is bounded, negative values of $\beta$
are not in principle ruled out (the same happens in spin systems).

In the case $\beta > 0$, we could interpret energy as a level of dissatisfaction for the
agents, as their spontaneous arrangement in a configuration with high $\mathcal{H}$ 
would be extremely improbable. Instead they would prefer to
arrange themselves in a configuration which minimizes $\mathcal{H}$.

\section{A Hamiltonian function for agents}

We can cast the Hamiltonian into a more friendly expression, by employing the more symmetric notation 

\begin{equation}
\left<a, b\right> = 2\delta(a, b)-1,
\end{equation}
such that $\left<a, b\right>$=1 if $a=b$, -1 otherwise. This is motivated by analogy with spins models. Renaming 

\begin{eqnarray}
J=\frac{\lambda_J}{\beta N(N-1)} \\
C=\frac{\lambda_C}{2 \beta N} \\
R=-\frac{\lambda_J+2\lambda_R}{\beta N(N-1)},
\label{eq_parameters}
\end{eqnarray}
we can write $\mathcal{H}$ compactly as

\begin{equation}
\mathcal{H} = -\frac{1}{2}J\sum_{i=1}^N \sum_{<j\neq
i>}\left<S_i, S_j\right> + \frac{1}{2}\sum_{i=1}^N \sum_{<j\neq i>}R
-C\sum_{i=1}^N \left<S_i, B_i\right>,
\label{eq_model}
\end{equation}
where $\sum_{<j \neq i>}$ indicates sum among all agents $j$ closer than $R_c$
to $i$. This expression now strikingly resembles a spin Hamiltonian. The first term represents the
energy associated with assortative ($J > 0$) or disassortative ($J < 0$) behavior, the second term
can be understood as the cost of overcrowding ($R > 0$) or as a preference towards agglomeration ($R < 0$). 
Finally the last term is the energy associated with the consequence among
agents, and we will always assume $C > 0$ (otherwise the agents have a tendency to have $S_i \neq B_i$, and this is ``unphysical'' in this context).

To understand the meaning of this Hamiltonian in a social sense, we can interpret it as encoding the following facts:

\begin{enumerate}
\item The external opinion does not necessarily agree with the internal opinion, and the cost of internal disagreement is proportional to $C$.
\item Two remote agents (further apart than a distance $R_c$) cannot influence each other in terms of opinions.
\item Agents who tend to agree with their peers have $J>0$ and the cost of
disagreeing with other agents is proportional to $|J|$. Similarly, agents who tend to disagree with
their peers have $J<0$, and the cost of agreeing with other agents is proportional to $|J|$.
\item For $R>0$, the cost of every pair of neighbors (closer than $R_c$) is proportional to $|R|$.
\end{enumerate}

Eq. \ref{eq_model} constitutes our model, and is the main result of this work.
The most evident difference with a spin Hamiltonian is the overcrowding term,
proportional to $R$. In lattice spin models with fixed number of neighbors,
this term amounts to a constant shift in energy and is therefore omitted.
Similarly, in the following analysis, in order to focus on the more relevant effects of 
$J$ and $C$, we will neglect this term (setting $R=0$) by assuming that $J$ is
much larger than $R$. Thus we are studying a particular slice of the phase space
$(J, C, R)$, namely $(J, C, R=0)$. This is not the same as neglecting the constraint on
$P(r_{ij} < R_c|H)$, as this would set $\lambda_R=0$ in Eq. \ref{eq_parameters}, 
and $R$ would be non-zero.

In order to complete the analogy with spin models, we can devise an order
parameter, similar to a magnetization. One possible definition is the \emph{decision} order parameter,

\begin{equation}
M = \frac{1}{N} \max\left(\sum_{i=1}^N \delta(S_i, q)\right), 1 \leq q \leq Q,
\end{equation}
which is just the average proportion of agents with the dominant opinion (if
any). Table I shows each parameter with their intended social meaning (and
thermodynamical analogy, if applicable).

\begin{table}[t]
\begin{center}
\begin{tabular}{lll}
\hline
\hline
Symbol & Social meaning & Thermodynamic analog \\
\hline
$\mathbf{r}_i$ & Position & Particle position \\
$S_i$ & External opinion & Particle spin orientation \\
$B_i$ & Internal opinion & Particle local field \\
\hline
$N$   & Number of agents & Number of particles \\
$Q$   & Number of possible opinions & Magnitude of spin \\
$R_c$ & Influence radius & Interaction cutoff \\
$C$   & Consequence level & Local anisotropy \\
$J$   & Social personality & Exchange coupling \\
$R$   & Cost of overcrowding & Repulsion energy \\
$T$   & Social temperature & Temperature \\
$E$   & Dissatisfaction & Internal energy \\
$M$   & Decision level & Magnetization \\
\hline
\hline
\end{tabular}
\end{center}
\caption{Parameters in our model, their meaning and thermodynamic analogs.}
\end{table}

\section{Results}

We can understand the phenomenology of this model by considering a few agents.
Intuitively, several situations which are familiar to us in real life will arise, 
depending on the value of $J/C$. For a given realization of $J/C < 1$, we can
imagine an individual for whom the cost of keeping his/her internal, private
opinion $B$ is to leave a certain group. On the other hand, the case $J/C > 1$
could be represented by an individual who in order to belong to a certain group,
has to hide his/her opinion $B$, expressing publicly a different opinion $S$. 
Also, ``peer pressure'' could be imagined as the case of an individual
surrounded by several other peers having different opinion: if he/she cannot
escape, then is coerced to hide his/her internal opinion and express the opinion
of the group.  This is similar to the rules of the ancient game of Go, where there are stones which are captured when 
surrounded by enemy stones. For instance, a single white stone when surrounded by four adjacent (non-diagonal) 
black stones is captured and turns black. In our model, this would be represented by a situation with 
$C < 4J$ (four bonds with different peers is enough to surpass the energy of consequence). Thus, it can be seen 
that according to the different possible values in parameter space, there exists a richness of cases which can be linked
to real life situations.

Some analytical results can be obtained in the limit $J \rightarrow 0$ (or,
equivalently, $R_c \rightarrow 0$). There the agents are completely 
non-interacting (because we have neglected the overcrowding parameter $R$) and we can solve the 
problem exactly. The internal energy is given by,

\begin{equation}
\frac{\left<\mathcal{H}\right>}{NC} = \frac{e^{-\beta C}(Q-1)-e^{\beta C}}{e^{\beta C}+e^{-\beta C}(Q-1)},
\end{equation}

and the system always presents a second-order phase
transition\cite{NoteTransitions} where $E(T)$ has
its inflection point at $\tau_c=T_c/C$, given by

\begin{equation}
\exp(2/\tau_c) = (1-Q)\left(\frac{\tau_c+1}{\tau_c-1}\right).
\label{eq_tauc_q}
\end{equation}

In order to further explore the implications of our model, we have used the Monte Carlo Metropolis 
algorithm~\cite{LandauBinder} to compute thermodynamical averages for the assortative case ($J > 0$) under fixed $T$,
$J$, and $C$ (i.e, following the canonical ensemble of Eq.
\ref{eq_probability}). In our numerical implementation, we considered $N=20$ agents
contained in a square box of side $L=50$ units. It is important to stress here that we are not concerned with the thermodynamic limit of a statistical mechanical
theory, as the MaxEnt formalism does not depend on a large number of degrees of
freedom ~\cite{Jaynes2003}. As the number of agents increases we of course
decrease the uncertainty of the predictions of the model (as $1/\sqrt{N}$), but
in principle, MaxEnt offers a valid prediction (the optimal prediction under
limited information) for any $N > 1$. Phase transitions exist and are well defined in
finite systems~\cite{Gross2000, Carignano2002} (also see the articles by Chomaz et
al~\cite{Chomaz1999, Chomaz2006} on the criteria for their definition). 
In the following, we use the term phase transition as applies for finite systems.

For each temperature, we equilibrated the system over the first 800 thousand MC steps, after which we accumulated averages over 1 million steps.

\begin{figure}
\vspace{20pt}
\begin{center}
\includegraphics[scale=0.17]{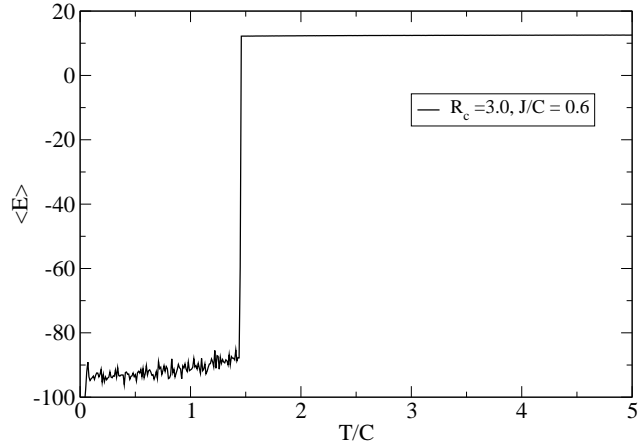}
\end{center}
\caption{Internal energy as a function of temperature, for a case in which a
first-order phase transition is observed. The values of the parameters are
$\rho=0.06$ and $J/C=0.6$.}
\label{fig_first_ord}
\end{figure}

\begin{figure}
\vspace{20pt}
\begin{center}
\includegraphics[scale=0.17]{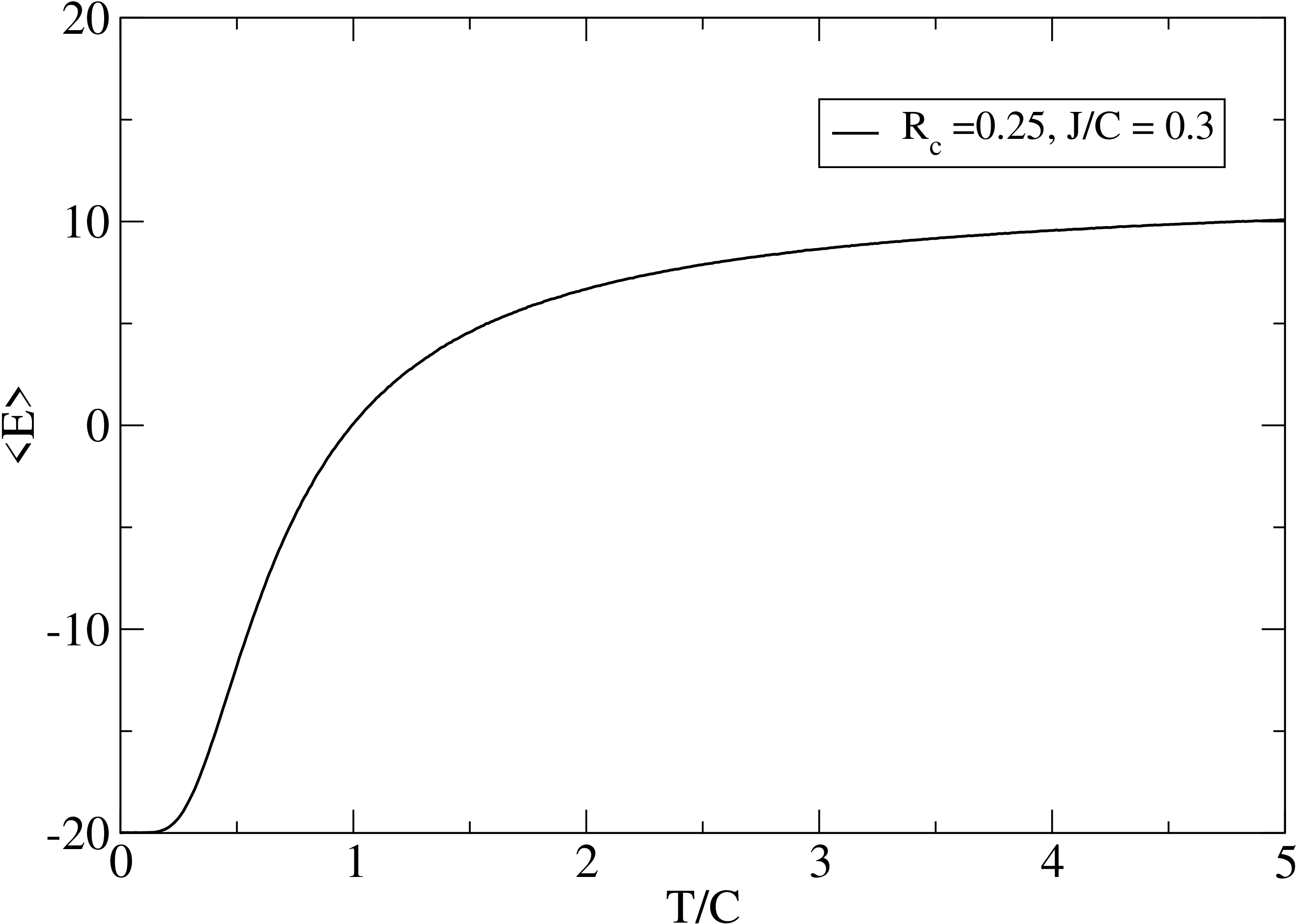}
\end{center}
\caption{Internal energy as a function of temperature, for a case in which a
second-order phase transition is observed. The values of the parameters are
$\rho=0.005$ and $J/C=0.3$.}
\label{fig_second_ord}
\end{figure}

We observe that, depending on the values of $\rho=R_c/L$ and $J/C$, an increase in temperature can induce either an abrupt, 
first-order phase transition (as shown in Fig. \ref{fig_first_ord}), or a smooth, second-order phase transition (as Fig. \ref{fig_second_ord} shows).

We recognize these two cases simply by detailed inspection of the curves. In the case 
of continuous phase transitions we could always find an intermediate temperature
for which the system is thermodynamically stable at an intermediate energy, and
this is not the case for the abrupt phase transitions. In the latter case, there
is always an energy gap between two phases. We would expect that microcanonical
simulations could produce states in between this gap, but this is outside the
scope of this work.

\begin{figure}
\vspace{20pt}
\begin{center}
\includegraphics[scale=0.17]{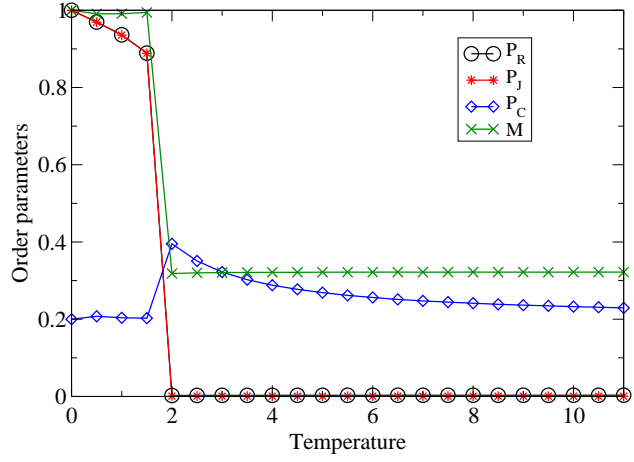}
\end{center}
\caption{Probabilities $P_R$, $P_J$, $P_C$ and the decision parameter $M$ as a function of temperature, 
for the first-order phase transition at $\rho=0.005$ and $J/C=0.3$.}
\label{fig_first_ord_M}
\end{figure}

In the case of a first-order phase transition, for $T$ above the transition temperature 
$T_c$, the system is completely disordered (see Fig. \ref{fig_first_ord_M}), both in terms of position (agents act like free particles in a gas) 
and opinion (no clearly marked preference), whereas below $T_c$ the agents are spatially clustered and have a single opinion.
We could interpret this kind of phase transition as a ``social breakdown'', a
state of dissociation between beliefs and opinions. Above $T_c$ individuals have lost all trace of
their personality, their internal beliefs and their interpersonal relations.

Note that the $E(T)$ curve is flat above $T_c$, which is expected because we have 
no kinetic degrees of freedom in the Hamiltonian, so there is an upper limit to
the amount of energy that can be stored in the system. In social terms this
means that, under this regime, after the transition individuals immediately reach their upper limit 
for dissatisfaction.

\begin{figure}
\vspace{20pt}
\begin{center}
\includegraphics[scale=0.17]{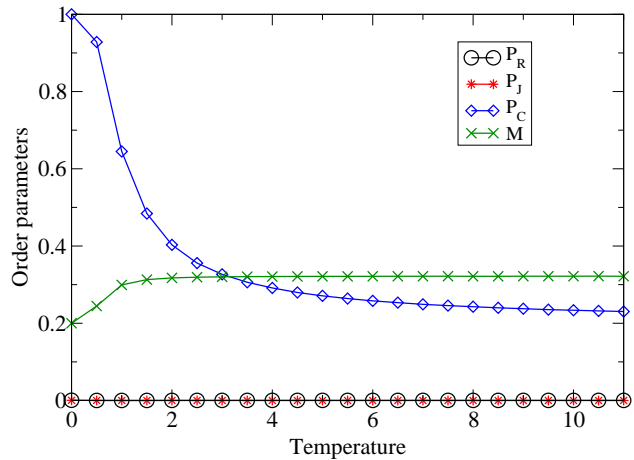}
\end{center}
\caption{Probabilities $P_R$, $P_J$ and $P_C$ and the decision parameter $M$ as a function of temperature, 
for the second-order phase transition at $\rho=0.06$ and $J/C=0.6$.}
\label{fig_second_ord_M}
\end{figure}

In the case of a second-order phase transition, the system is always spatially disordered 
(see Fig. \ref{fig_second_ord_M}), because the interaction $J$ that tends to 
bind them is much weaker than the internal consequence $C$. As the temperature increases, 
the individuals gradually lose their internal consequence not as a collective, but one by one. In this regime, the social
system demonstrates a tolerance to the diversity of opinions and the ``social breakdown'' is softened.

\begin{figure}
\vspace{20pt}
\begin{center}
\includegraphics[scale=0.17]{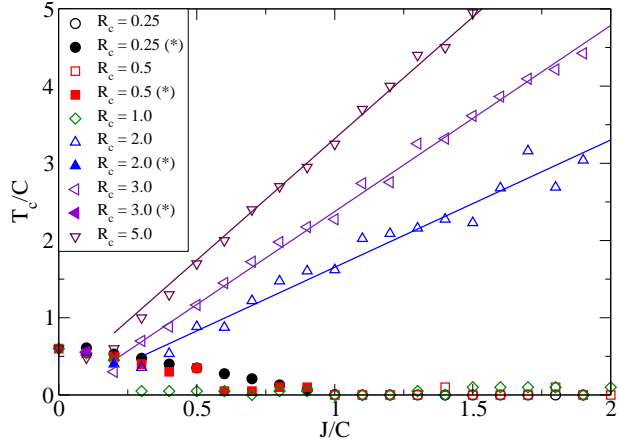}
\end{center}
\caption{Reduced transition temperature $\tau_c=T_c/C$ as a function of $\gamma=J/C$ for
different values of $\rho$. Curves marked with (*) in the legend are
second-order phase transitions.}
\label{fig_phasediag_rho}
\end{figure}

\begin{figure}
\vspace{20pt}
\begin{center}
\includegraphics[scale=0.17]{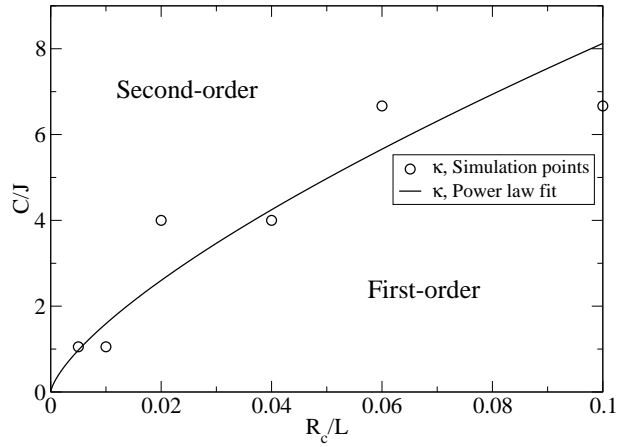}
\end{center}
\caption{The critical value of $C/J$, $\kappa$, as a function of $\rho=R_c/L$.}
\label{fig_kappa}
\end{figure}

Fig. \ref{fig_phasediag_rho} shows the transition temperature 
$\tau_c=T_c/C$ for different values of $J/C$ and $\rho$, presenting both first-order and
second-order phase transitions. There is a critical value $C/J=\kappa$, such 
that for $C > \kappa J$ the transition is second-order, and for $C \leq \kappa J$ 
the transition is first-order. Fig. \ref{fig_kappa} shows an estimation of $\kappa$ as a function 
of the interaction radius $\rho$.

\begin{figure}
\vspace{20pt}
\begin{center}
\includegraphics[scale=0.17]{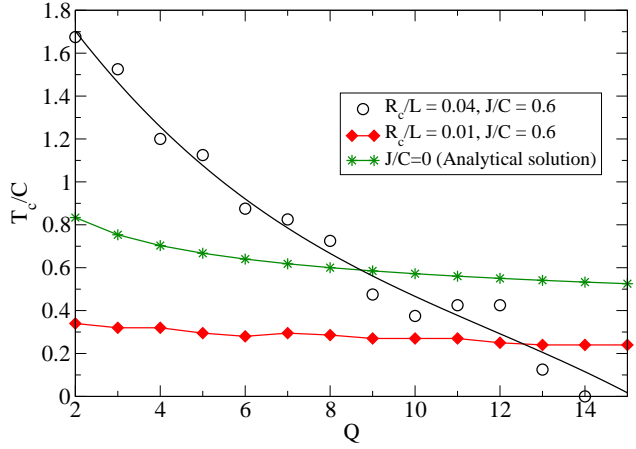}
\end{center}
\caption{Reduced transition temperature $\tau_c$ as a function of $Q$. Open circles
correspond to first-order phase transitions, filled diamonds to
second-order. Stars correspond to equation \ref{eq_tauc_q}.}
\label{fig_phasediag_q}
\end{figure}

Fig. \ref{fig_phasediag_q} shows the dependence of the transition temperature
with increasing $Q$, the number of possible beliefs or opinions. As expected,
$Q$ contributes to the entropy associated with opinions and thus decreases the
transition temperature. This can be understood as revealing the following fact: 
diversity of choices always tends to quicken the onset of the breaking of
uniformity, regardless of whether social interaction or personal beliefs are more prevalent.

\section{Discussion}

We have found that the minimum noticeable level of internal consequence $C$ is
given by $\kappa J$, a fact which we interpret as follows: the consequence of an
individual must overcome a critical value, corresponding to the amount of
``opposing'' individuals surrounding him/her needed to change his/her own opinion.

When the level of internal consequence falls below this critical value, the
individuals have effectively a null internal consequence. In this case, they will 
either ``follow the masses'' (at low temperatures) or ``follow no one'' (at high 
temperatures). The transition between these two phases is abrupt, first-order, which
means there is no state of coexistence, no state where some people are
``mass-followers'' and some are not. We can imagine this catastrophic change as
the dominance of random behavior in agents, where suddenly the whole population loses empathy
(connection with others) and rationality (they start disregarding their own
internal beliefs).

On the other hand, when $C$ overcomes this critical value, we begin to see
continuous, second-order phase transitions. The presence of continuous transitions 
reveals a system undergoing a smooth change in state (from a fully convinced population 
at low temperatures to fully undecided individuals at high temperatures). Thus, when 
$C$ has a greater influence over $J$, the agents' behavior tends to be governed
mostly by their own beliefs instead of the influence (or ``social pressure'')
exerted by close people. This causes the individual opinion to persist, and therefore, we can 
argue that each agent is reasonably in agreement with his vision of the world. In 
thermodynamical terms, this means that the internal energy gradually increases as the 
imposed temperature increases (the system has a finite, positive heat capacity at all energies).

\section{Conclusions}

We have proposed a Hamiltonian model, deduced from maximum entropy statistical
inference, for a society where individuals can be described only by a personal, 
internal opinion (or belief) and an external opinion (or behavior). Each individual interacts 
with others in his/her immediate surroundings and can alter his/her own behavior as a function of the social 
group he/she belongs to. Our model produces several kinds of agent distributions, 
which we qualitatively associate with real-life organizations and social phenomena. It is
important to note that the derivation of our model relies only on known values
for the probabilities $P(S_i = B_i|H)$ and $P(S_i = S_j|(r_{ij} < R_c) \wedge H)$, which can be 
measured directly in a real-life social setting. The associated parameters $J$, $C$, $R$ or
the temperature can be inferred from these probabilities (Eqs.
\ref{eq_multi_PJ}, \ref{eq_multi_PC}, \ref{eq_multi_PR} and \ref{eq_beta}).

Based on our numerical results, we conclude that for an homogeneous society there 
is a particular ``critical'' line in parameter space, namely $C=\kappa J$, defining a 
lower limit for measurable consequence, below which the internal belief is completely 
forgotten and any given individual only acts as a function of others. 

\begin{acknowledgments}
We thank fruitful discussions with Diego Contreras. SD acknowledges financial support from Fondecyt grant 3110017.
\end{acknowledgments}


\begin{thebibliography}{29}
\bibitem[{Not()}]{NoteTransitions}
\bibinfo{note}{We classify phase transitions as follows. A second-order (or
  continuous) phase transition is characterized by an inflection point in the
  $E(T)$ curve, while a first-order (or abrupt) phase transition is signaled by
  a discontinuous jump in the $E(T)$ curve.}
\end{thebibliography}

\end{document}